\documentclass[a4,11pt]{article} 
\usepackage{amssymb}
\usepackage{amsmath}
\usepackage[pdftex]{graphicx}
\usepackage{cite}
\usepackage{slashed}
\usepackage{bm}
\usepackage{fancyhdr}
\usepackage{float}
\usepackage{tocloft}
\usepackage[top=2.5cm, bottom=2.0cm, left=2.5cm, right=2cm]{geometry}

\begin{document}

\begin{center}
{\large \bf	The cross-section for the $\gamma e^{-}  \rightarrow Ze^{-} \rightarrow l^{-} l^{+} e^{-} $ scattering at the LHeC }\\

\vspace*{1cm}

 { Bui Thi Ha Giang$^{a,}$  \footnote{giangbth@hnue.edu.vn}}\\

\vspace*{0.5cm}
 $^a$ Hanoi National University of Education, 136 Xuan Thuy, Hanoi, Vietnam
\end{center}

\begin{abstract}
A measurement of Z production cross-section in $\gamma e^{-}$ collision at Large Hadron-electron Collider (LHeC) is presented to compare to that at International Linear Collider (ILC). The total cross-section depends strongly on the polarization of the initial and final $e^{-}$ beams, the electron beam energy $E_{e}$ with the energy of the proton beam taken to be $E_{p} = 7$ TeV. The result shows that the total cross-section in $\gamma e^{-} \rightarrow Z e^{-} \rightarrow l^{-}l^{+}e^{-}$ at LHeC is much larger than that at ILC.\\

\end{abstract}

\section{Introduction}
\hspace*{1cm} The electro-weak and strong interactions have been unified in the Standard model (SM). The discovery of Higgs signal at the Large Hadron Collider (LHC) has completed the particle spectrum of the SM. However, the existence of some theoretical drawbacks in the SM has motivated the extended models. One of the most attractive extended models is Randall-Sundrum (RS) model which provides a new scenario with the extra dimension to solve the gauge hierarchy naturally \cite{rs}. Due to the extra dimension, the additional scalar field called radion can be mixed with Higgs boson \cite{jung, eboos, frank, ali}. The low mass radion region, even 10 GeV, is worth investigating \cite{ahm}. Besides the SM couplings, the anomalous couplings of the mixed radion and Higgs boson to $\gamma \gamma, ZZ, gg, \gamma Z$ have been considered . \\
\hspace*{1cm} It is important to test the electro-weak sector of the SM through studying the Z boson production. The Z production using the Compton backscattered photon was considered at International Linear Collider (ILC) \cite{cif}. Recently, Z-pair production in hadron colliders has been investigated at the LHC \cite{ste, gon, wang, hein, cms18,raf,raha}. However, the Large Hadron-electron Collider (LHeC), planned at the Large Hadron Collider (LHC), is a cleaner configuration \cite{yanbi, lhec1, bri, lhec2, lhec3}. Moreover, the center of mass energy at the LHeC is higher than that at the ILC. The incoming proton beam energy at LHeC has been $E_{p} = 7$ TeV and the center of mass energy has been $\sqrt{s} = 2\sqrt{E_{p}E_{e}}$. The LHeC has planned to collide electrons with energy from 60 GeV to 140 GeV \cite{rod}. Therefore, it can provide a better condition for researching a lot of new phenomena compared to the LHC and ILC. The LHeC may play a significant role in the pursuit of new physics beyond the SM \cite{akay}. $\gamma e^{-}$ collision with the photon beam radiated from the proton has provided an extra experimental scenario to help reduce the background \cite{yue}. In spite of a lower luminosity, $\gamma e^{-}$ subprocess can be studied as a complementary tool to $e^{-} p$ collision at the LHeC. \\
\hspace*{1cm} In our present work, we have studied the subprocess $\gamma e^{-} \rightarrow Ze^{-} \rightarrow l^{-} l^{+} e^{-}$, including the vertices of Z boson as, $\gamma ZZ$, $\gamma\gamma Z$,  $\gamma Zh$, $\gamma Z \phi$. With the contribution of new interactions in RS model, including radion, Higgs propagators, the total cross-section has been expected to detect in experiments. The layout of this paper is as follows. In Section II, we review the Randall-Sundrum model and the mixing of Higgs - radion. The cross-section for the $\gamma e^{-} \rightarrow Ze^{-} \rightarrow l^{-} l^{+} e^{-}$ collision at the LHeC is presented in Section III. Finally, we summarize our results and make conclusions in Section IV.
\section{A review of Randall-Sundrum model and the mixing of Higgs-radion}
\hspace*{1cm}The RS model consists of one extra dimension bounded by two 3-branes cite{ahm}. The UV-brane, which the fifth dimension is bounded at, located at $y = 0$ and the IR-brane located at $y = \pi r_{c}$. The five dimensional metric has the following form:
\begin{equation}
ds^{2} = e^{-2kb_{0}|y|}\eta_{\mu\nu}dx^{\mu}dx^{\nu} - b_{0}^{2}dy^{2},
\end{equation}
where $b_{0}$ is a length parameter for the fifth dimension, $r_{c}$ is the compactification radius and $k$ is the curvature of the five dimensional geometry. The exponential represents the warp factor which generates the gauge hierarchy. The values of the bare parameters are determined by the Planck scale and the applicable value for size of the extra dimension is assessed by $kr_{c} \simeq 11 - 12$. Thus the weak and the gravity scales can be naturally generated. The relation $\overline{M}_{Pl}^{2} = M_{5}^{3}/k$ is derived from the 5D action. The scale of physical phenomena on the IR-brane is given by $\Lambda_{\phi} \equiv \overline{M}_{Pl} e^{-kr_{c}\pi}$ with $\Lambda _{\phi } \sim $ few TeV. The mass of the $n^{th}$ KK graviton excitation is $m^{G}_{n} = x_{n}^{G}k\Lambda_{\phi}/\overline{M}_{Pl}$, with $x_{n}^{G}$ being the roots of the $J_{1}$ Bessel function \cite{rue}. The coupling strength of the graviton KK states to the SM fields is $\Lambda_{\phi}^{-1}$. The gravity-matter interactions have been obtained \cite{kow} 
\begin{equation}
\mathcal{L}_{int} = -\dfrac{1}{\hat{\Lambda}_{W}}\Sigma_{n\neq 0} h^{n}_{\mu\nu} T^{\mu\nu} - \dfrac{\phi_{0}}{\Lambda_{\phi}} T^{\mu}_{\mu},
\end{equation}
where $h^{n}_{\mu\nu} (x)$ are the Kaluza-Klein (KK) modes of the graviton field $h_{\mu \nu}(x, y)$, $\phi_{0} (x)$ is the radion field, $\Lambda_{\phi} \equiv \sqrt{6}M_{Pl}\Omega_{0}$ is the VEV of the radion field and $\hat{\Lambda}_{W} \equiv \sqrt{2}M_{Pl}\Omega_{0}$. The $T^{\mu\nu}$ is the energy-momentum tensor, which is given at the tree level \cite{bae, csa}
\begin{equation}
T^{\mu}_{\mu}=\Sigma_{f} m_{f} \overline{f}f - 2m^{2}_{W}W^{+}_{\mu}W^{-\mu}-m^{2}_{Z}Z_{\mu}Z^{\mu} + (2m^{2}_{h_{0}}h_{0}^{2} - \partial_{\mu}h_{0}\partial^{\mu}h_{0}) + ...
\end{equation}
The gravity-scalar mixing is described by the following action\cite{domi}
\begin{equation}
S_{\xi } =\xi \int d^{4}x \sqrt{g_{vis} } R(g_{vis} )\hat{H}^{+} \hat{H},
\end{equation}
where $\xi $ is the mixing parameter, $R(g_{vis})$ is the Ricci scalar for the metric $g_{vis}^{\mu \nu } =\Omega _{b}^{2} (x)(\eta ^{\mu \nu } +\varepsilon h^{\mu \nu } )$ induced on the visible brane, $\Omega _{b} (x) = e^{-kr_{c} \pi} (1 + \frac{\phi_{0}}{\Lambda _{\phi }})$ is the warp factor, $\hat{H}$ is the Higgs field in the 5D context before rescaling to canonical normalization on the brane.
 With $\xi \ne 0$, there is neither a pure Higgs boson nor pure radion mass eigenstate. This $\xi$ term mixes the $h_{0}$ and $\phi_{0}$ into the mass eigenstates $h$ and $\phi$ as given by 
\begin{equation} 
\left(\begin{array}{c} {h_{0} } \\ {\phi _{0} } \end{array}\right)=\left(\begin{array}
{cc} {1} & {6\xi \gamma /Z} \\ {0} & {-1/Z} \end{array}\right)\left(\begin{array}{cc}
 {\cos \theta } & {\sin \theta } \\ {-\sin \theta } & {\cos \theta } \end{array}\right)
 \left(\begin{array}{c} {h} \\ {\phi } \end{array}\right)=\left(\begin{array}{cc}
  {d} & {c} \\ {b} & {a} \end{array}\right)\left(\begin{array}{c} {h} \\ {\phi } \end{array}\right), \label{pt}
\end{equation}
where
$Z^{2} = 1 + 6\gamma ^{2} \xi \left(1 -\, \, 6\xi \right) = \beta - 36\xi ^{2}\gamma ^{2}$ is the coefficient of the radion kinetic term after undoing the kinetic mixing, $\gamma = \upsilon /\Lambda _{\phi }, \upsilon = 246$ GeV, $a = -\dfrac{cos\theta}{Z}, b = \dfrac{sin\theta}{Z}, c = sin\theta + \dfrac{6\xi\gamma}{Z}cos\theta, d = cos\theta - \dfrac{6\xi\gamma}{Z}sin\theta$. The mixing angle $\theta $ is
\begin{equation}
\tan 2{\theta } = 12{\gamma \xi Z}\frac{m_{h_{0}}^{2}}{m_{\phi _{0}}^{2} - m_{h_{0}}^{2} \left( Z^{2} - 36\xi^{2} \gamma ^{2} \right)},
\end{equation}
where $m_{h_{0}}$ and $m_{\phi _{0}}$ are the Higgs and radion masses before mixing.\\
The new physical fields h and $\phi $ in (\ref{pt}) are Higgs-dominated state and radion, respectively
\begin{equation} 
m_{h,\phi }^{2} =\frac{1}{2Z^{2} } \left[m_{\phi _{0} }^{2} +\beta m_{h_{0} }^{2} \pm \sqrt{(m_{\phi _{0} }^{2} +\beta m_{h_{0} }^{2} )^{2} -4Z^{2} m_{\phi _{0} }^{2} m_{h_{0} }^{2} } \right].
\end{equation}
\\
 There are four independent parameters $\Lambda _{\phi } ,\, \, m_{h} ,\, \, m_{\phi } ,\, \, \xi$ that must be specified to fix the state mixing parameters. We consider the case of $\Lambda _{\phi } = 5$ TeV and $\frac{m_{0} }{M_{P} } = 0.1$, which makes the radion stabilization model most natural \cite{csa}.

\section{The cross-section for the $\gamma e^{-} \rightarrow Ze^{-} \rightarrow l^{-} l^{+} e^{-}$ collision}
\begin{figure}[!htb] 
\begin{center}
\includegraphics[width= 14 cm,height= 10 cm]{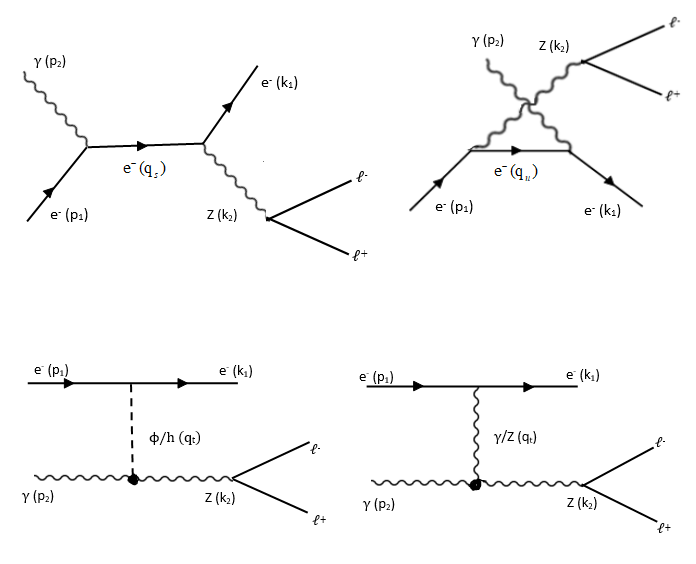}
\caption{\label{feynman} Feynman diagrams for $\gamma e^{-} \rightarrow Ze^{-} \rightarrow l^{-} l^{+} e^{-}$ collision, representing the s, u, t-channels, respectively.}
\end{center}
\end{figure}

\begin{figure}[!htb] 
\begin{center}
\includegraphics[width= 8 cm,height= 5 cm]{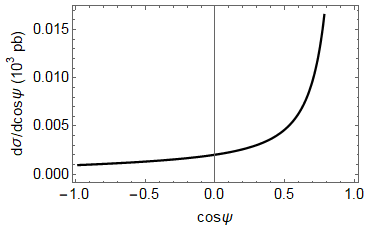}
\caption{\label{Fig.1} The differential cross-section with respect to the $cos\psi$. The parameters are chosen as $P_{1} = P_{2} =  0.8$, $E_{e}$ = 60 GeV. }
\end{center}
\end{figure}

\begin{figure}[!htb] 
\begin{center}
\includegraphics[width= 8 cm,height= 5 cm]{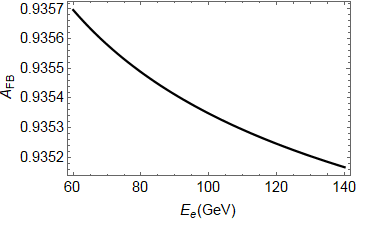}
\caption{\label{Fig.2} The forward-backward asymmetry with respect to the electron beam energy $E_{e}$. The parameters are taken to be $P_{1} = P_{2} =  0.8$. }
\end{center}
\end{figure}

\begin{figure}[!htb] 
\begin{center}
\includegraphics[width= 8 cm,height= 6 cm]{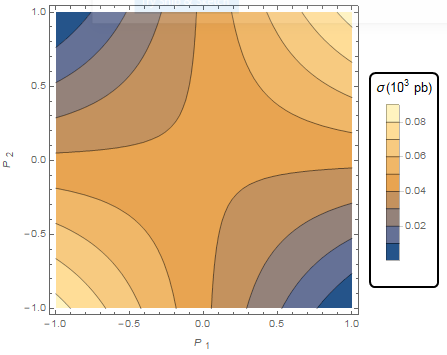}
\caption{\label{Fig.3} The total cross-section as a function of the polarization coefficients of initial and final electron beam. The electron beam energy is chosen as $E_{e} = 60$ GeV. }
\end{center}
\end{figure}

\begin{figure}[!htb] 
\begin{center}
\includegraphics[width= 8 cm,height= 5.5 cm]{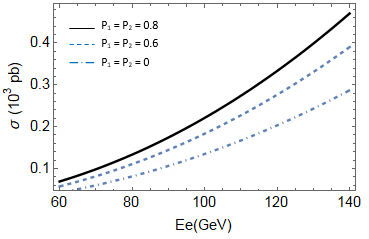}
\caption{\label{Fig.4} The total cross-sections as a function of the electron beam energy $E_{e}$ in the cases of the polarization coefficients $P_{1} = P_{2} = 0.8, 0.6, 0$.}
\end{center}
\end{figure}

\hspace*{1cm} Feynman rules for the couplings in the RS model are shown as follows \cite{ahm}
\begin{align}
&g_{eeh} = -i\overline{g}_{eeh} = -i\dfrac{gm_{e}}{2m_{W}}\left( d + \gamma b\right),\\
&g_{ee\phi} = -i\overline{g}_{ee\phi} = -i\dfrac{gm_{e}}{2m_{W}}\left( c + \gamma a\right),\\
&C_{\gamma Zh} = \dfrac{\alpha}{2\pi\nu_{0}}\left[2g_{h}^{r}\left(\dfrac{b_{2}}{tan \theta_{W}} - b_{Y}tan \theta_{W}\right)-g_{h}\left(A_{F} + A_{W}\right)\right],\\
&C_{\gamma Z\phi} = \dfrac{\alpha}{2\pi\nu_{0}}\left[2g_{\phi}^{r}\left(\dfrac{b_{2}}{tan \theta_{W}} - b_{Y}tan \theta_{W}\right)-g_{\phi}\left(A_{F} + A_{W}\right)\right].
\end{align}
where a, b, c, d are the state mixing parameters in RS model. $g_{h} = d + \gamma b $, $g_{\phi} = c + \gamma a $, $g_{h}^{r} = \gamma b$, $g_{\phi}^{r} = \gamma a$, the triangle loop functions   $A_{F}, A_{W}$ are given in Ref.\cite{gun} \\
\hspace*{1cm}We consider the collision process in which the initial state contains electron and photon which emitted from proton beams, the final state contains Z boson and electron 
\begin{equation} \label{pt1}
 e^{-} (p_{1}) + \gamma (k_{2}) \rightarrow e^{-} (k_{1}) + Z (k_{2}).
\end{equation}
Here, $p_{i}, k_{i}$ (i = 1, 2) stand for the momentums. There are three Feynman diagrams contributing to reaction (\ref{pt1}), representing the s, u, t-channel exchange depicted in Fig.\ref{feynman}.\\
\\
\hspace*{1cm}The transition amplitude representing the s-channel is given by
\begin{equation}
M_{s} =  -i\dfrac{e\overline{g}_{eZ}}{q^{2}_{s} - m^{2}_{e}}\varepsilon^{*}_{\nu} (k_{2})\gamma^{\nu} \left(v_{e}-a_{e}\gamma^{5}\right)\overline{u}(k_{1}) \left(\slashed{q}_{s} + m_{e}\right) \varepsilon_{\mu} (p_{2})\gamma^{\mu}u(p_{1}).
\end{equation}
\hspace*{1cm}The transition amplitude representing the u-channel can be written as
\begin{equation}
M_{u} =-i\dfrac{e\overline{g}_{eZ}}{q^{2}_{u} - m^{2}_{e}} \overline{u}(k_{1}) \gamma^{\mu}\varepsilon_{\mu}(p_{2})\left(\slashed{q}_{u} + m_{e}\right)\varepsilon^{*}_{\nu} (k_{2})\gamma^{\nu} \left(v_{e}-a_{e}\gamma^{5}\right)  u(p_{1}) . 
\end{equation}
\hspace*{1cm}The transition amplitude representing the t-channel is given by
\begin{equation}
M_{t} = M_{Z} + M_{\gamma} + M_{\phi} + M_{h},
\end{equation}
here
\begin{align}
&M_{\gamma} = -\dfrac{e}{q^{2}_{t} }\overline{u}(k_{1})\gamma^{\beta}u(p_{1})\eta_{\sigma\beta} \varepsilon^{*}_{\nu} (k_{2})\Gamma^{\mu\sigma\nu}_{\gamma \gamma Z}(p_{2}q_{t} k_{2}) \varepsilon_{\mu} (p_{2}),\\
&M_{Z} = -\dfrac{\overline{g}_{eZ}}{q^{2}_{t} - m^{2}_{Z}}\overline{u}(k_{1})\gamma^{\beta}\left(v_{e}-a_{e}\gamma^{5}\right)u(p_{1})\left(\eta_{\sigma\beta} - \dfrac{q_{t\sigma}q_{t\beta}}{m^{2}_{Z}}\right)\varepsilon^{*}_{\nu} (k_{2})\Gamma^{\mu\sigma\nu}_{\gamma ZZ}(p_{2}q_{t} k_{2})\varepsilon_{\mu} (p_{2}),\\
&M_{\phi} = i\dfrac{\overline{g}_{ee\phi}C_{\gamma Z \phi}}{q^{2}_{t} - m^{2}_{\phi}}\overline{u}(k_{1})u(p_{1})\varepsilon^{*}_{\nu} (k_{2})\left(p_{2}k_{2}\eta^{\mu\nu} - p_{2}^{\nu}k_{2}^{\mu}\right)\varepsilon_{\mu} (p_{2}),\\
&M_{h} = i\dfrac{\overline{g}_{eeh}C_{\gamma Zh}}{q^{2}_{t} - m^{2}_{h}}\overline{u}(k_{1})u(p_{1})\varepsilon^{*}_{\nu} (k_{2})\left(p_{2}k_{2}\eta^{\mu\nu} - p_{2}^{\nu}k_{2}^{\mu}\right)\varepsilon_{\mu} (p_{2}).
\end{align}
\hspace*{1cm}The triple gauge boson couplings are given by \cite{raha}
\begin{equation} 
\Gamma^{\sigma\mu\nu}_{\gamma \gamma Z}(p_{2}q_{t}k_{2}) =\dfrac{g_{e}}{m_{Z}^{2}} p_{2}p_{2} \Biggl[h_{1}^{\gamma}(q_{t}^{\mu}\eta^{\sigma\nu}-q_{t}^{\nu}\eta^{\sigma\mu}) - h_{3}^{\gamma}\varepsilon^{\mu\sigma\nu\alpha} q_{t\alpha} \Biggr],
\end{equation}
\begin{equation} 
\Gamma^{\sigma\mu\nu}_{\gamma ZZ}(p_{2}q_{t}k_{2}) = -\dfrac{g_{e}}{m_{Z}^{2}} p_{2}p_{2} \Biggl[f_{4}^{\gamma}(p_{2}^{\nu}\eta^{\mu\sigma} + p_{2}^{\sigma}\eta^{\mu\nu}) - f_{5}^{\gamma}\varepsilon^{\mu\sigma\nu\alpha} (q_{t\alpha} - k_{2\alpha}) \Biggr],
\end{equation}
The total cross-section for the whole process can be calculated as follows 
\begin{equation}
\sigma = \sigma (\gamma e^{-} \rightarrow  Z e^{-}) \times Br(Z\rightarrow l^{-}l^{+}).
\end{equation}
The effective production cross-section for the subprocess at the LHeC can be described \cite{yue}
\begin{equation}
\sigma (\gamma e^{-} \rightarrow  Z e^{-}) = \int^{\xi_{max}}_{Max(\dfrac{(m_{e} + m_{Z})^{2}}{s}, \xi_{min)}}E_{p}f(\xi E_{p})d\xi\int^{(cos\psi)_{max}}_{(cos\psi)_{min}}\dfrac{d\widehat{\sigma}(\widehat{s})}{dcos\psi}dcos\psi,
\end{equation} 
where
\begin{equation}
\frac{d{\widehat{\sigma}(\widehat{s})}}{d(cos\psi)} = \frac{1}{32 \pi \widehat{s}} \frac{|\overrightarrow{k}_{1}|}{|\overrightarrow{p}_{1}|} |M_{fi}|^{2}
\end{equation}
is the expressions of the differential cross-section \cite{pes}, $\widehat{s} = 4E_{e}E_{\gamma} = \xi s$ is center of mass energy of the sub-process $\gamma e^{-} \rightarrow  Z e^{-}$. $\psi = (\widehat{\overrightarrow{p}_{1}, \overrightarrow{k}_{2}})$ is the scattering angle. The photon flux can be written as
\begin{equation}
f(\xi E_{p}) = \int_{Q^{2}_{min}}^{Q^{2}_{max}}\dfrac{dN_{\gamma}}{dE_{\gamma}dQ^{2}}dQ^{2},
\end{equation}
where
\begin{equation}
\dfrac{dN_{\gamma}}{dE_{\gamma}dQ^{2}} = \dfrac{\alpha_{e}}{\pi}\dfrac{1}{E_{\gamma}Q^{2}}\left[\left(1-\dfrac{E_{\gamma}}{E_{p}}\right)\left(1-\dfrac{Q^{2}_{min}}{Q^{2}} \right)F_{E}+\dfrac{E^{2}_{\gamma}}{2E^{2}_{p}}F_{M}\right],
\end{equation}
with 
\begin{align}
&Q^{2}_{min} = \dfrac{M^{2}_{P}E^{2}_{\gamma}}{E_{p}(E_{p}-E_{\gamma})},\\
&F_{E} = \dfrac{4M^{2}_{P}G^{2}_{E} + Q^{2}G_{M}^{2}}{4M^{2}_{P} + Q^{2}},\\
&G^{2}_{E} = \dfrac{G^{2}_{M}}{\mu^{2}_{P}} = \left(1+\dfrac{Q^{2}}{Q^{2}_{0}} \right)^{-4},\\
&F_{M} =  G^{2}_{M}.
\end{align}
$M_{P}$ is the mass of the proton. $E_{p}$ is the energy of the incoming proton beam. $E_{\gamma} = \xi E_{p}$ is the photon energy, which is related to the loss energy of the emitted proton beam. $F_{E}, F_{M}$ are functions of the electric and magnetic form factors given in the dipole approximation. $\mu^{2}_{P} = 7.78$ is the magnetic moment of the proton. \\
\hspace*{1cm}For numerical calculations, we choose the energy of the incoming proton beam at LHeC $E_{p} = 7$ TeV, the electron beam energy $E_{e}$ in the range 60 GeV $ \leq E_{e} \leq 140$ GeV \cite{rod}. The vacuum expectation value (VEV) of the radion field is $\Lambda_{\phi} = 5$ (TeV) \cite{domi}. The radion mass has been selected $m_{\phi} = 10$ GeV \cite{soa1}. The Higgs mass $m_{h} = 125$ GeV (CMS). The maximum value of anomalous couplings in the tightest limits with the corresponding observable are chosen as $f_{4}^{\gamma} = 2.4 \times 10^{-3}$, $f_{5}^{\gamma} = 2.7 \times 10^{-3}$, $h_{1}^{\gamma} = 3.6 \times 10^{-3}$, $h_{3}^{\gamma} = 1.3 \times 10^{-3}$ \cite{raha}. The polarization coefficients of initial and final $e^{-}$ beams are taken to be $P_{1} = P_{2} = 0.8$, respectively \cite{bri}. Since the contribution to the above integral formula is very small for $Q^{2}_{max} > 2 GeV^{2}$, therefore $Q^{2}_{max}$ is chosen as 2 $GeV^{2}$ \cite{yue}. We give estimates for the cross-sections as follows \\
\hspace*{1cm}i) The electron beam energy is chosen as $E_{e} = 60$ GeV. In Fig.\ref{Fig.1}, we plot the differential cross-section as a function of the $cos\psi$. From the figure, we can see that $d\sigma/dcos\psi$ is peaked in the forward direction and flat in the backward direction.\\
\hspace*{1cm}ii) In Fig.\ref{Fig.2}, the forward-backward asymmetry is plotted as a function of $E_{e}$. The figure indicates that the forward-backward asymmetry decreases when the electron beam energy increases. \\
\hspace*{1cm}iii) In Fig.\ref{Fig.3}, we evaluate the dependence of the total cross-section on the initial and final electron beam in the case of $E_{e} = 60$ GeV. The cross-section achieves the maximum value when both initial and final electron beams are left or right polarized $(P_{1} = P_{2} = \pm 1)$.  The cross-section has the minimum value when $P_{1} = 1, P_{2} = -1$ and vice versa. It is worth noting that the cross-section in the case of both left or right polarized initial and final electron beams is twice as much as the unpolarized electron beams.  \\
\hspace*{1cm}iv) In Fig.\ref{Fig.4}, we evaluate the dependence of the total cross-section on the electron beam energy $E_{e}$. The electron beam energy is chosen in the range of 60 GeV$ \leq \sqrt{s} \leq$ 140 GeV. Here, the polarization coefficients of initial and final electron beam are chosen as $P_{1} = P_{2} = 0.8, 0.6, 0$. The figure shows that the total cross-sections increase when the electron beam energy $\sqrt{s}$ increases. In Table \ref{tab1}, some typical values for the total cross-section are measured. We emphasize that the main contribution to the description process is by the s, u-channels, while the anomalous couplings for the contribution are very small. However, the study of anomalous couplings concerning the mixing of Higgs-radion needs to be further studied.

\begin{table}[!htb]
\centering
\caption{\label{tab1}Some typical values for the total cross-section in the $\gamma e^{-} \rightarrow Z e^{-} \rightarrow l^{-}l^{+}e^{-}$ collisions at the LHeC.} 

\begin{tabular}{|c|c|c|c|c|c|c|c|c|c|} 
\hline 
$E_{e}$ (GeV) & 60&70&80&90&100&110&120&130&140 \\ 
\hline 
$\sigma$($P_{1} = P_{2} = 0.8$) ($10^{3} pb$) &0.069&0.098&0.133&0.174&0.221&0.274&0.333&0.399&0.471\\
\hline
$\sigma$($P_{1} = P_{2} = 0.6$) ($10^{3} pb$) &0.057 &0.081 &0.110 &0.144 &0.183 &0.227&0.276&0.331&0.390 \\
\hline
$\sigma$($P_{1} = P_{2} = 0$) ($10^{3} pb$) &0.042 &0.059&0.081&0.106&0.135&0.167&0.203&0.243&0.287\\
\hline
\end{tabular} 
\end{table}
\newpage
\section{Conclusion}
\hspace*{1cm} In this work, we have evaluated the total cross-section in $\gamma e^{-} \rightarrow Z e^{-} \rightarrow l^{-}l^{+}e^{-}$ production at LHeC in the RS model. The result shows that the main contribution to the description process is by the s, u-channels, while the anomalous couplings for the contribution are very small. Note that the Z boson with such a short lifetime is never seen directly in the experiments. Even though the branching ratio of $Z \rightarrow l^{-}l^{+}$ is fairly small, the $l^{-}l^{+}$ channel leaves a relatively clean signature in the detector, which allows the process to be measured with a higher accuracy \cite{tana}. In this study, our interest is comparison between the cross-section at ILC and at LHeC. The value at ILC has been read in detail in Ref.\cite{cif}. The results that the total cross-section in $\gamma e^{-} \rightarrow Z e^{-} \rightarrow l^{-}l^{+}e^{-}$ at LHeC is much larger than that at ILC.\\
\hspace*{1cm} Finally, we emphasize that the scalar anomalous concerning the mixing of Higgs- radion in the RS model needs to be investigated further.\\
{\bf Acknowledgements}: The work is supported in part by Hanoi National University of Education under Grant No. SPHN21 – 07.\\


\end{document}